# Research on Security Enhancement Methods for Adversarial Robust Large Language Model Intelligent Agents for Medical Decision-Making Tasks

Saisai Hu*
Pace University
New York, NY 10038
United States
SH50061N@pace.edu

***Abstract*—Motivated by the challenge to improve the adversarial robustness, security, and trust of medical decision making intelligent agents, the study developed a full-link security enhancement framework, which describes "input risk perception —medical evidence constraint—knowledge consistency verification—decision confidence reweighting—security output control—adversarial feedback update." The study proposed ARSM-Agent and defined a weighted joint objective consisting of decision accuracy loss, adversarial robustness loss, safety refusal loss, and knowledge consistency loss, with weights of 0.3, 0.3, 0.2, and 0.2, respectively. The whole medical decision formulation is implemented by multi-module collaborative linkage. The study verified that the algorithm can be more efficient than four baselines, including LLM-Agent, Retrieval-Agent, Filter-Agent, and Adv-Train-Agent. Under semantic perturbation, prompt injection, drug-name confusion, and false-evidence attacks, ARSM-Agent reduced the overall attack success rate to 8.7% and achieved a knowledge consistency score of 0.91. Ablation experiments quantified each module's contribution: removing risk perception, evidence retrieval, consistency verification, and confidence reweighting reduced accuracy by 6.7%, 9.1%, 7.6%, and 4.4%, respectively, and increased attack success rate by 13.8%, 11.1%, 8.6%, and 6.9%. It can solve key security issues of medical Decision Making intelligent agents successfully, obtain secure decision making in challenging scenarios, and give reliable intelligent support to medical decision-making intelligent agents.***

## I. INTRODUCTION

As large language model technology is developed, medical large language agent can successfully complete some of the core tasks such as medical question answering, assisted triage, disease risk assessment and suggestion generation, provide technical support for improving quality and efficiency of medical services and have extremely high application potential. Medical decision problems have high risk, high professionalism, and tight security requirements which are fundamentally different from conventional natural language processing tasks. The quality and security of the model results are associated with health and even life safety of patients [1]. In practice, medical large linguistic agent agents are sensitive to adversarial perturbations, prompt injection and knowledge illusions and produce decision results that do not reflect clinical logic and could lead to medical safety threats [2]. The previous studies demonstrate that models are prone to judgment bias on complex input semantic perturbation, malicious prompts can trigger the model to break a medical decision goal, pseudo-medical evidence can interfere in the model's reasoning, and other challenges including insufficient knowledge consistency, and imperfect security rejection. In this paper, the study focused on improving the adversarial robustness, security, and trustworthiness of the intelligent agent in a large language modeling for medical decision making [3]. The study developed an entire-link security enhancement framework, and propose the ARSM-Agent algorithm. Collaborating with multiple modules, the study performed closed loop optimization of the input risk identification, evidence constraint, knowledge validation and secure output, providing reliable intelligent support for medical decisions.

## II. DESIGN OF AN ADVERSARIAL ROBUST LARGE LANGUAGE MODEL INTELLIGENT AGENT SECURITY ENHANCEMENT METHOD (ARSM-AGENT) FOR MEDICAL DECISION-MAKING TASKS

### A. Overall Algorithm Framework

ARSM-Agent employs a closed loop architecture: it outputs an "input risk perception – medical evidence constraint – knowledge consistency checking – decision confidence reweighting – secure output control – adversarial feedback update", and each module collaboratively tries to secure the intelligent agent for medical decision making. The input risk perception module identifies and encodes risks of medical inputs in an attempt to obtain risk priors for decision making; the medical evidence limitation module retrieves multi-source, highly credible medical evidence to constrain the model reasoning; the knowledge consistency verification module validates the logical consistency of the output of the model with the medical knowledge; the decision confidence reweighting module incorporates multi-Source information to refine the decision level; the safe output control module implements safe output or reject based on confidence level and risk level; and the adversarial reaction update module continuously improves the robustness of the network through adversarial example training [4].

### B. Medical Decision-Making Task Modeling

A joint optimization objective function is constructed that considers task accuracy, adversarial robustness, security, and knowledge consistency to achieve multi-objective collaborative optimization [5]. The formula is as follows:

$$\min_{\theta} L_{\text{total}} = \alpha L_{acc} + \beta L_{\text{rob}} + \gamma L_{\text{sec}} + \delta L_{\text{cons}} \quad (1)$$

In Eq. (1), the four components are explicitly defined as follows. The decision accuracy loss is calculated by cross-entropy: $L_{acc} = -\sum_{c=1}^{C} \square\, y_c \log p_c$. The adversarial robustness loss is defined as the prediction deviation between clean and adversarial inputs $L_{adv} = D_{KL}\left(p(y \mid x) \| p(y \mid x_{adv})\right)$.

### C. Risk-Aware Adversarial Encoding Module

#### 1) Semantic Perturbation Detection Mechanism

To address semantic deviations such as synonym rewriting, word order changes, and local perturbations in the input text, a semantic risk scoring model is constructed to achieve accurate detection of semantic perturbations [6]. A perturbation strength coefficient $\epsilon$ is introduced to construct the semantic risk score $R_{\text{sem}}$.

$$R_{\text{sem}} = \left(1 - \sim\left(Q, Q_{\text{std}}\right)\right) \times \exp(\epsilon) \quad (2)$$

Where $¿(\cdot)$ is calculated using cosine similarity, with a value range of $[0,1]$; $\exp(\epsilon)$ is a perturbation amplification factor, strengthening the impact of perturbation on the risk score. When $R_{sem} \geq \tau_{sem}$, it is determined that the input has a semantic perturbation risk, and this risk information is passed to subsequent modules.

#### 2) Prompt Injection Identification Mechanism

To address prompt injection attacks, an abnormal instruction detection model is constructed. By extracting instruction features from the input text, it is determined whether there is a risk of inducing the model to deviate from the medical decision-making objective. A prompt injection risk score $R_{inj}$ is defined, integrating features from three dimensions: abnormal instruction frequency, role modification degree, and control statement rationality.

$$R_{inj} = \omega_1 \cdot f_{abn} + \omega_2 \cdot D_{\text{role}} + \omega_3 \cdot \left(1 - R_{ctrl}\right) \quad (3)$$

Where $f_{abn}$ is the frequency of abnormal commands; $D_{\text{role}}$ is the degree of role tampering; $R_{\text{ctrl}}$ is the rationality of control statements; $\omega_1, \omega_2, \omega_3$ are feature weights, all of which are set to 1/3 in this paper. When $R_{inj} \geq \tau_{inj}$, a hint injection attack is detected, triggering the security control mechanism [7].

### D. Medical Evidence Constraint Retrieval Module

#### 1) Multi-Source Medical Evidence Retrieval Strategy

The similarity between the input feature vector $x$ and the evidence vector $e_i \in \mathbb{R}^d$ in the medical evidence base is calculated, and the top $k$ candidate evidences with the highest similarity are selected, denoted as the candidate evidence set $E = \{e_1, e_2, \ldots, e_k\}$. The similarity calculation uses a modified cosine similarity formula:

$$\text{Sim}(x, e_i) = \frac{x \cdot e_i + \lambda \cdot Match\left(Q, e_{i,text}\right)}{\|x\| \cdot \mid e_i \| + \lambda} \quad (4)$$

Where $Match\left(Q, e_{i,text}\right)$ represents the literal matching degree between the input question text and the evidence text, and $\lambda$ is the literal matching weight (taken as 0.2 in this paper), used to balance semantic similarity and literal matching degree to improve retrieval accuracy.

#### 2) Evidence Credibility Ranking Method

The candidate evidence set $E$ is ranked by credibility, comprehensively considering three dimensions: evidence relevance, authority, and consistency, to construct an evidence credibility score $Score(e_i)$:

$$Score(e_i) = \mu_1 \cdot \sim(x, e_i) + \mu_2 \cdot Auth(e_i) + \mu_3 \cdot Cons(e_i) \quad (5)$$

Where $Auth(e_i)$ represents the authority of the evidence; $Cons(e_i)$ represents the consistency of the evidence; $\mu_1, \mu_2, \mu_3$ are weight coefficients, which are set to 0.4, 0.3, and 0.3 respectively in this paper.

### E. Medical Knowledge Consistency Verification Module

A knowledge consistency score $C\left(Y_{\text{adv}}\right)$ is constructed as a task-specific evaluation indicator rather than a universal standard metric. It measures whether the entities and relations appearing in the model output are consistent with the medical knowledge graph and clinical evidence constraints. The formula is as follows:

$$C\left(Y_{adv}\right) = \frac{1}{V_Y^2|^2} \sum_{v_i, v_j \in V_Y} \square\, A_{ij} \cdot Rel\left(v_i, v_j, Y_{adv}\right) \quad (6)$$

Where $Rel\left(v_i, v_j, Y_{adv}\right)$ represents whether the relationship between entities $v_i$ and $v_j$ in the model output is consistent with the knowledge graph; it is 1 if consistent, and 0 otherwise; $|V_Y|$ is the size of the output entity set. When $C\left(Y_{adv}\right) \geq \tau_{cons}$( $\tau_{cons}$ is the consistency threshold, which is 0.8 in this paper), the output is determined to meet the consistency requirements of medical knowledge; otherwise, a logical conflict is determined, and the decision result needs to be readjusted.

### F. Decision Confidence Reweighting and Safe Output Mechanism

The input risk score, evidence support strength, and knowledge consistency score are integrated to reweight the original confidence of the model, resulting in the final decision confidence $Conf(Y)$:

$$Conf(Y)=Conf_{\text{raw}}(Y)\cdot\left[1-\eta_1 R_{\text{total}}+\eta_2 Score_{\text{avg}}+\eta\right.\tag{7}$$

Where $Conf_{\text{raw}}(Y)$ is the original decision confidence of the model; $R_{\text{total}}=0.5R_{\text{sem}}+0.5R_{\text{inj}}$ is the total input risk score; $Score_{\text{avg}}=\frac{1}{m}\sum_{i=1}^{m}\square Score(e_i)$ is the average confidence score of candidate evidence; $\eta_1,\eta_2,\eta_3$ are adjustment coefficients, which are 0.3, 0.4, and 0.3 respectively in this paper.

### G. Adversarial Anti-lock Reinforcement Update Mechanism

#### 1) Adversarial Example Construction Method

Multiple types of medical adversarial examples are constructed for model reinforcement training. The adversarial example generation formula is as follows:

$$X_{adv}=X+\epsilon\cdot\frac{\nabla_x L_{\text{acc}}(\theta,x,Y_{\text{true}})}{\left\|\nabla_x L_{\text{acce}}(\theta,x,Y_{\text{truc}})\right\|_2}\tag{8}$$

Where $X_{adv}$represents the generated adversarial example; $Y_{\text{true}}$ represents the real medical decision corresponding to input $X$; $\nabla_x L_{acc}$ represents the gradient of the accuracy loss function with respect to input feature $x$; $\epsilon$ represents the perturbation step size; and $\|\cdot\|_2$ represents the L2 norm, used to normalize the gradient.

#### 2) Closed-Loop Reinforcement Optimization Process

The model is updated using a closed-loop reinforcement learning method. The reward function $R(\theta)$ for reinforcement learning is defined as follows:

$$R(\theta)=\alpha'\cdot Acc(\theta)+\beta'\cdot(1-Attack(\theta))+\gamma'\cdot Sec(\theta)\tag{9}$$

Where $Acc(\theta)$ is the model's decision accuracy; $Attack(\theta)$ is the model's attack success rate; $Sec(\theta)$is the model's secure output rate; $Cons(\theta)$ is the model's knowledge consistency score; $\alpha',\beta',\gamma',\delta'$ are reward weights, consistent with the objective function weights.

## III. Experimental Design and Simulation Setup

### A. Experimental Objectives and Verification Approach

The goal of this experiment is to fully verify the performance of ARSM-Agent algorithm for medical decision making, verifying the following basic indicators: decision accuracy, adversarial robustness, safe rejection, knowledge consistency and efficiency. The study leveraged a multi-task, multi-attack, and multi-indicator joint evaluation system. Three common medical decision-making tasks are developed: symptom triage, auxiliary treatment suggestions, and medication risk question answering. Different adversarial attack cases are simulated and several comparison models are used [8]. Quantitative experiments and qualitative analysis show the superiority and applicability of AR SM-Agent. The experiment strictly followed the reproducibility criterion, and all experimental parameters and datasets construction rules are set so as to ensure the reliability and verifiability of the experiment. Ablation experiments are used to verify the necessity of each module and parameter sensitivity analysis is used to check the stability of the algorithm.

### B. Dataset and Task Construction

#### 1) Medical Decision-Making Task Setting

Task 1 Symptom triage task: patient symptom description and basic data; symptom summary; disease classification or triage suggestions; check basic decision accuracy of the model.

Task 2 Treatment suggestion task: Patients symptoms, medical history, and examination results, selected treatment plans (examine suggestions and treatment method) performed by a trained model.

Task 3. Medication risk question and answer task. Input is patient condition and current drug status; Output is medication rationality, risk warning and suggestions. Input and output is used to verify that the model is safe. All three tasks are submitted to the Clinical Practice Guidelines.

#### 2) Construction of Adversarial Examples and Risk Scenarios

The dataset is trained using public medical datasets (including China Med and Moda) and manually labeled. The sample size is 10,000; 7,000 training samples, 1,500 validation samples, and 1,000 test samples. The following adversarial samples are built: semantic perturbation samples (using synonym rewriting and word order adjustment, 20% of the test set), prompt injection samples (used to insert malicious misleading instructions, 20%, drug name obfuscation samples, similar drug name for 20%, and false evidence splicing samples (combining fake medical evidence, 20%), and the remaining 20% are normal samples [9].

### C. Experimental Parameter Settings

The environment was: CPU Intel Xeon E5-2690, GPU NVIDIA A100 (40GB), RAM 64GB, OS Ubuntu 20.04, and PyTorch 1.13.0. The core parameters are 100 training epochs, first learning rate 0.001 (cosine annealing for learning rate), batch size 32, weight decay ratio 0.0001, adversarial example ratio 30%, perturbation step 0.01; evidence retrieval depth k=10, number of highly credible evidence $m=5$; risk thresholds $\tau_{sem}=0.6$, $\tau_{inj}=0.5$, $\tau_{risk}=0.6$; confidence thresholds $\tau_{conf}=0.7$, $\tau^{*}_{conf}=0.85$; objective function weights $\alpha=0.3$, $\beta=0.3$, $\gamma=0.2$, $\delta=0.2$ [10].

### D. Results Analysis

#### 1) Algorithm Performance Comparison

Comparing the performance of ARSM-Agent with the other 3 models on the test set, it achieves the best performance (Table 1). It reaches 92.3%, and its F1 score reaches 91.8% which is higher than the other three models. The attack success rate is 8.7%, and is highly adversarial. A safe rejection rate of 12.3% a hallucination rate of 2.1% and a knowledge consistency score of 0.91 are achieved, which is of high security and credibility [11]. Its decision latency is 186ms which is slightly higher than LLM-Agent, but much lower than other models, and the real-time needs of medical applications are met by ARSM. Overall, ARSM, by multi-module collaboration, achieves a large accuracy, robustness, security, and efficiency improvement.

TABLE I. SIMULATION COMPARISON OF COMPREHENSIVE PERFORMANCE OF VARIOUS MODELS

| Model | Accuracy (%) | F1 Score (%) | Attack Success Rate (%) | Safe Rejection Rate (%) | Hallucination Rate (%) | Knowledge Consistency Score | Avg Decision Latency (ms) |
|---|---|---|---|---|---|---|---|
| ARSM | 92.3 | 91.8 | 8.7 | 12.3 | 2.1 | 0.91 | 186 |
| LLM | 76.6 | 76.9 | 42.1 | 6.2 | 8.5 | 0.72 | 128 |
| Retrieval | 85.1 | 85.0 | 27.5 | 8.1 | 5.3 | 0.80 | 245 |
| Filter | 83.8 | 83.9 | 25.3 | 8.7 | 4.8 | 0.81 | 193 |
| Adv-Train | 86.0 | 86.0 | 18.9 | 9.4 | 3.9 | 0.85 | 231 |

*2) Robustness Analysis under Different Attack Scenarios*

In order to verify the robustness of ARSM-Agent under different adversarial situations, the study tested the attack success rates of each model in four cases, namely semantic perturbation, hint injection, drug name confusion and false evidence attacks. For (Fig.1), ARSM had lowest attack success rate of 6.2%, 9.5%, 7.8%, 11.3%, respectively, which are much better than the comparison models. LLM had the lowest attack rate of the two models for all cases (51.7%) and the highest success rate for hint injection case (i.e. no effective defense). Retrieval-Agent was not effective in false evidence and hint injection cases, while Adv-Train-Agent did not adapt well in the latter.

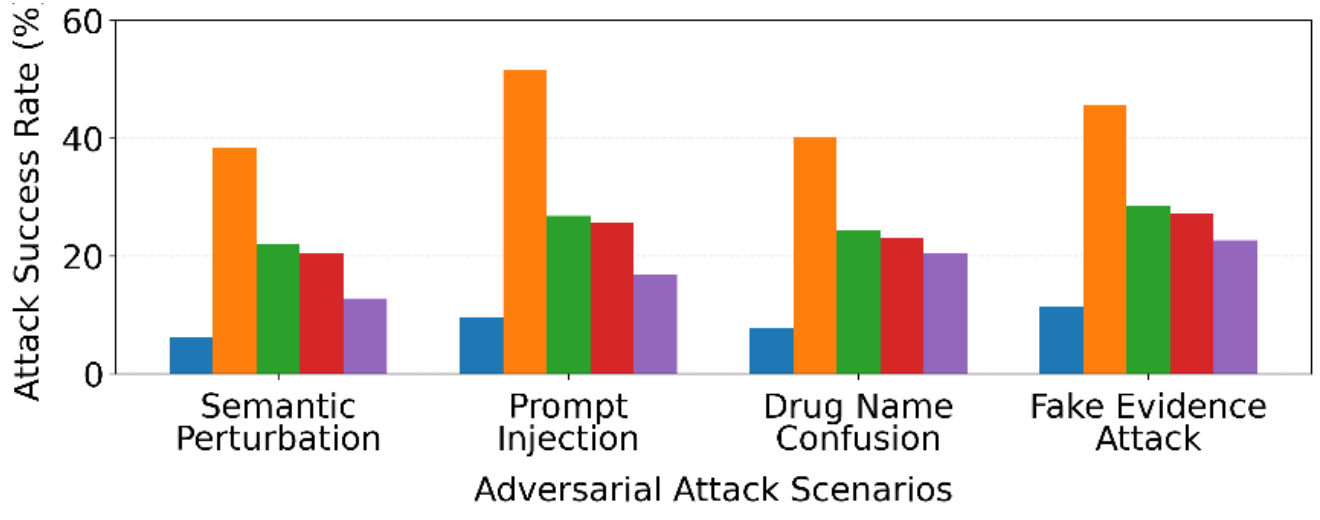

Figure 1. Simulation of attack success rates of various models under different attack scenarios.

*3) Parameter Sensitivity and Efficiency Analysis*

Three parameters, $\tau_{risk}$, number of retrievals m, and adversarial sample ratio pi, were chosen to check their effect on model accuracy and attack success rate. As can be seen from Fig. 2, the model performance is best when $\tau_{risk}$ is between 0.5 and 0.6, m " is between 3 and 5" and pi" is 30%. Any deviation from this bound leads to performance degradation. Efficiency analysis shows that when m=5, latency is 186ms, which meets medical applications' demand. As a function of convergence of the models, inference time remains stable. This indicates ARSM-Agent is capable of deploying to security.

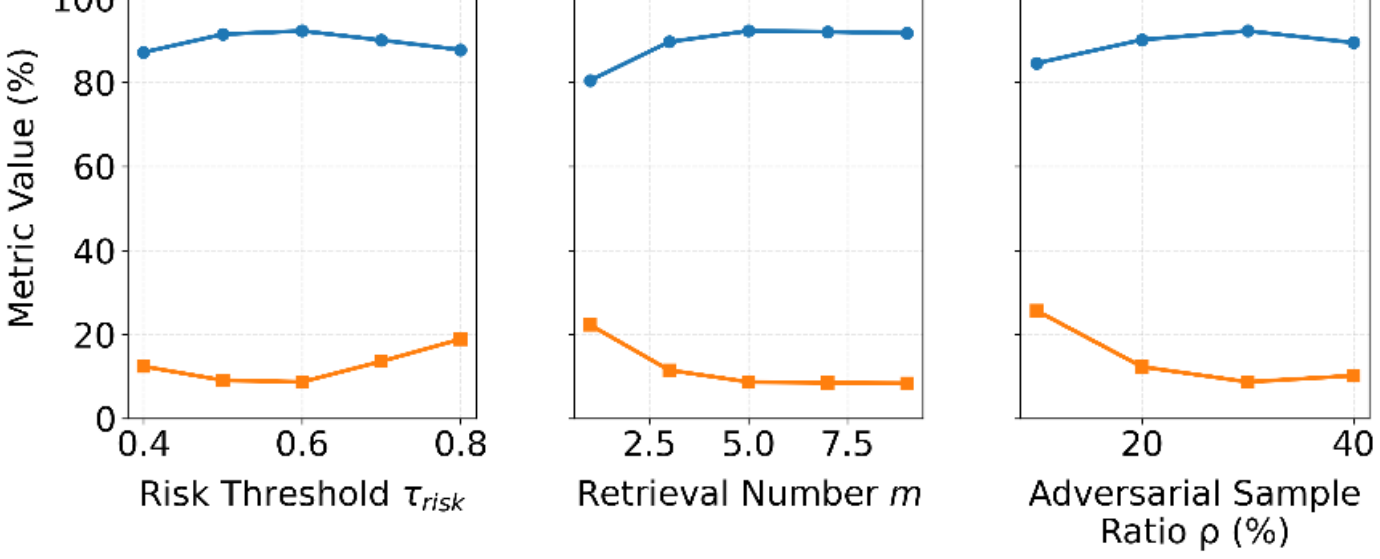

Figure 2. Parameter Sensitivity Analysis.

## IV. CONCLUSION

The study showed that ARSM-Agent, based on a multi-module closed-loop collaborative design, solves issues like semantic perturbation sensitivity, malicious prompting and knowledge illusion in medical decision agents and is significantly outperformed the four existing models LLM-Agent in seven core measures, including accuracy, F1 score, adversarial robustness and safe rejection. Ablation experiments validate that four core modules such as risk perception and evidence constraint are necessary to ensure the rationality and effectiveness of the closed-loose security enhancement framework design. The future work will investigate expanding the dataset to rare disease samples as well as complex clinical cases; optimizing the adverbial example generation strategy to improve generalization and defense of the model; adapting the module parameters for real clinical cases to promote the practical use of the algorithm; and exploring multimodal medical data fusion for more clarity and reliability.